\documentclass[12pt]{article}
\usepackage[dvips]{graphicx}
\usepackage[dvips]{epsfig}

\date{}

\begin{document}

\begin{center}

\textbf{ON THE POSSIBILITY OF FERROMAGNETISM AND HALF-METALLICITY IN LOCAL MOMENT SYSTEMS}

\vspace{1cm}

L. HARITHA, G. GANGADHAR REDDY\\
\textit{Department of Physics, Kakatiya University, Warangal-506 009, India}\\

\vspace{0.4cm}

A.RAMAKANTH\\
\textit{Stanley College of Engineering and Technology for Women, Hyderabad-500 001, India}\\
 
\end{center}  

\vspace{0.5cm}

\begin{center}
\textbf{Abstract}
\end{center}

We use the Kondo lattice model to investigate the possibility of ferromagnetism and half-metallicity in local moment systems. Using the spectral density approach and making use of the fact that the   spontaneous magnetization of local moment and the itinerant electron polarization are coupled, we derive an expression for the paramagnetic susceptibility. The magnetic ordering temperature is determined from the singularities of the susceptibility. The magnetic phase diagram is constructed in $T-n$(band filling) plane. It is found that ferromagnetism is possible only for small values of $n$. It is also found that the temperature drives the transition of the system from half-metal to metal.

\vspace{1.0cm}


\vspace{0.5cm}

\newpage

\section {Introduction}

\noindent Local moment systems constitute one of the most fascinating fields in both theory and experiment in condensed matter physics$^{1,2}$.  In general there is a consensus that local moments play a decisive role in a variety of phenomenon such as magnetism, heavy fermions, high temperature superconductivity, colossal magnetoresistance and spintronics. Many of the real systems that are classified as local moment systems consist of both the localized and itinerant electrons. In the case of rare earth compounds and alloys, for example, the electrons in the partially filled 4f-shell are highly localized and maximize the magnetic moment according to Hund's rule and thus constitute the localized moment.  The electrons in 5d6s-bands constitute the itinerant electrons. Other prominent groups of materials whose electronic structure somewhat resembles that of rare-earth materials are the 3d transition metal oxides $^{3}$ like $La_{1-x}Sr_{x}MnO_{3}$ or $La_{1-x}Ca_{x}MnO_{3}$ and magnetic semiconductors $^{4,5}$ such as $Ga(Mn)As$.  Based on the spin-resolved photoemission studies $^{3}$ it is identified that the manganese perovskite $La_{0.7}Sr_{0.3}MnO_{3}$ is a half-metallic system well below the Curie temperature. The similarity to the rare-earth compounds stems from the fact that the 3d-band is split by the crystal field into the doubly degenerate $e_g$- and the triply degenerate $t_{2g}$-bands. The $t_{2g}$-band is occupied by three electrons with their spins aligned parallel due to the strong Hund's rule coupling and this provides the localized  spin back-ground with $S=3/2$ in which the itinerant $e_g$-band electrons move.  These electrons again couple ferromagnetically via Hund's exchange coupling  with the localized spins. 

Theoretically, the basic model for understanding the magnetic phenomena in systems where local magnetic moments couple ferromagnetically to itinerant carriers is the Kondo lattice model(KLM). The KLM in the strong coupling limit is also known as the double exchange model. Since the many-body problem of the KLM is not exactly solvable, several approximation schemes such as the coherent potential approximation$^{6}$, the second-order perturbation theory$^{7}$, the moment conserving decoupling approach $^{8}$, the dynamic mean-field theory$^{9-12}$ and the interpolation self-energy ansatz$^{13}$ have been developed. Even though all these schemes start from different limiting cases of the model, the conclusions drawn from them show remarkable similarities. 
In addition to these approximate methods, spectral density approach (SDA) has also proven to be very convenient to study various many-body problems. SDA is a reliable approximation in the strong coupling limit so long as the one-electron spectral density has predominantly a two-peak structure. In the literature SDA has been successfully used for the investigation of both the attractive$^{14}$ and the repulsive$^{15}$ Hubbard model, the periodic Anderson model$^{16}$ and the $t-J$ model$^{17}$.  Very recently this approach was also used to investigate$^{18}$ the interacting spin waves in the KLM. In this work the KLM is mapped on to an effective Heisenberg model and then SDA is applied. In addition, SDA was also applied to repulsive Hubbard model extended by Jahn-Teller interaction, to study$^{19}$ the interplay between magnetism and structural distortion in strongly correlated electron systems.

In the present work, we attempt an approximate solution of the KLM using the SDA.  The aim of the present investigation is to study the conditions under which the interband exchange and band filling together cause a collective (ferromagnetic) ordering and half-metallicity of the local-moment system. 

The present investigation is organized as follows. In the next section we introduce KLM and discuss the approximation used to solve the underlying many-body problem using SDA. To determine the magnetic ordering temperature, we derive the expression for static magnetic susceptibility in the third section. The fourth section is for a discussion and interpretation of the results. In the last section we conclude our results.

\section {Model and theory}
\noindent Magnetic materials whose magnetic properties depend on a system of localized electrons which are coupled via exchange interaction to itinerant conduction electrons are described by the KLM, which is given by 
\begin{equation}
H=H_{s}+H_{sf}
\end{equation}

\noindent $H_{s}$ describes the itinerant electrons in the presence of a homogeneous magnetic field $B$.

\begin{equation}
H_{s}=\sum_{i,j,\sigma}\left(T_{ij}-z_{\sigma}\mu_{B}B\delta_{ij}\right)c_{i\sigma}^{\dagger}c_{j\sigma}
 =\sum_{\mathbf{k},\sigma}\left(\epsilon_{\mathbf{k}}-z_{\sigma}\mu_{B}B\right)c_{\mathbf{k}\sigma}^{\dagger}c_{\mathbf{k}\sigma}
\end{equation}
$T_{ij}$ are the hopping integrals being connected by Fourier transformation to the free Bloch energies $\epsilon_{\mathbf{k}}$.
$\mu_{B}$ is the Bohr magneton. $c_{i\sigma}^{\dagger}(c_{j\sigma})$ and $c_{\mathbf{k}\sigma}^{\dagger}(c_{\mathbf{k}\sigma})$ are the creation(annihilation) operators of itinerant electron with spin $\sigma$ $(\sigma = \uparrow , \downarrow )$ at lattice site $\mathbf{R}_i$ and with a wavevector $\mathbf{k}$.  $z_\sigma$ is a sign factor, $z_{\sigma}=\delta_{\sigma
\uparrow}-\delta_{\sigma\downarrow}$. The second term in Eq.(1) is the interband exchange term with coupling strength $J$, written as intra-atomic interaction between the conduction electron spin $\sigma_i$ and the localized magnetic moment represented by the spin $\mathbf{S_i}$.

\begin{equation}
H_{sf}=-J\sum_{j}\mathbf{\sigma}_{j}\cdot \mathbf{S}_{j}
\end{equation}
It is more  convenient to express this exchange interaction in the following second quantized form:

\begin{equation}
H_{sf}=-\frac{1}{2}J\sum_{j,\sigma}(z_{\sigma}S_{j}^{z}c_{j\sigma}^{\dagger}c_{j\sigma}+S_{j}^{-\sigma}c_{j-\sigma}^{\dagger}c_{j\sigma})
\end{equation}
Where $S_{j}^{\sigma}=S_{j}^{x}+iz_{\sigma}S_{j}^{y}$.

We are mainly interested in the itinerant electron properties. So the relevant quantities to be calculated are the single-electron Greens functions and the spectral function.
\begin{equation}
G_{\mathbf{k}\sigma}(E)=\left<\left< c_{\mathbf{k}\sigma};c_{\mathbf{k}\sigma}^{\dag}\right>\right>_{E}
\hspace{0.5cm};\hspace{0.5cm}S_{\mathbf{k}\sigma}(E)=-\frac{1}{\pi}\Im G_{\mathbf{k}\sigma}(E+i0^{+})
\end{equation}
In the atomic limit and for classical spin, this Greens function has two poles at $\pm JS/2$.  Therefore it is realistic to 
make a two pole ansatz for spectral density in the finite band width case also:
\begin{equation}
S_{\mathbf{k}\sigma}(E)=\sum_{j=1}^{2}\alpha_{\sigma}(\mathbf{k},j)\delta\left(E-E_{\sigma}(\mathbf{k},j)\right)
\end{equation}
Here the spectral weights $\alpha_{\sigma}(\mathbf{k},j)$ and the poles $E_{\sigma}(\mathbf{k},j)$ are unknown
and are to be determined. For this, we use two definitions of spectral moments:
\begin{equation}
P_{\mathbf{k}\sigma}^{(m)}=\int dE\ E^{m}S_{\mathbf{k}\sigma}(E)
\end{equation}
and
\begin{equation}
P_{\mathbf{k}\sigma}^{(m)}=<[\ \underbrace{\left[\cdots\left[c_{\mathbf{k}\sigma},H\right]_{-},
\cdots,H\right]_{-}}_{m-fold},c_{\mathbf{k}\sigma}^{\dag}]_{+}>
\end{equation}
The two-peak ansatz for the spectral density involves only four unknowns. Therefore it is sufficient to calculate only
four moments. Calculating the commutators in Eq.(8) and using Eq.(6) in Eq.(7) we get the set of equations to 
determine the spectral weights $\alpha_{\sigma}(\mathbf{k},j)$ and the poles $E_{\sigma}(\mathbf{k},j)$ as
 follows:
$$\alpha_{\sigma}(\mathbf{k},1) +\alpha_{\sigma}(\mathbf{k},2)=1$$
$$E_{\sigma}(\mathbf{k},1)\alpha_{\sigma}(\mathbf{k},1) +E_{\sigma}(\mathbf{k},2)\alpha_{\sigma}(\mathbf{k},2)=
\epsilon_{\mathbf{k}\sigma}-z_{\sigma}\frac{J}{2}M$$
\begin{equation}
E_{\sigma}^{2}(\mathbf{k},1)\alpha_{\sigma}(\mathbf{k},1) +E_{\sigma}^{2}(\mathbf{k},2)\alpha_{\sigma}
(\mathbf{k},2)=\epsilon_{\mathbf{k}\sigma}^{2}-z_{\sigma}J\epsilon_{\mathbf{k}\sigma}M+\frac{J^{2}}{4}\left<S^{2}\right>
\end{equation}
$$E_{\sigma}^{3}(\mathbf{k},1)\alpha_{\sigma}(\mathbf{k},1) +E_{\sigma}^{3}(\mathbf{k},2)\alpha_{\sigma}
(\mathbf{k},2) 
= \\ \epsilon_{\mathbf{k}\sigma}^{3}-z_{\sigma}\frac{3}{2}J\epsilon^{2}_{\mathbf{k}\sigma}M+\frac{J^{2}}{2}\epsilon_{\mathbf{k}\sigma}
\left[\left<S^{2}\right>+\frac{M^{2}}{2}\right]-z_{\sigma}\frac{J^{3}}{8}M\left<S^{2}\right>$$\\
\noindent With  $M = <S^{z}>$ and $\epsilon_{\mathbf{k}\sigma} = \epsilon_{\mathbf{k}}-z_{\sigma}\mu_{B}B$. Solving the above set of equations we obtain

\begin{equation}
\alpha_{\sigma}(\mathbf{k},j)=\frac{1}{2}\left[1+(-1)^{j}\frac{(\epsilon_{\mathbf{k}\sigma}-z_{\sigma}JM)}
{\sqrt{\epsilon_{\mathbf{k}\sigma}^{2}-2z_{\sigma}J\epsilon_{\mathbf{k}\sigma}M+J^{2}\left<S^{2}\right>}}\right]
\end{equation}

\begin{equation}
E_{\sigma}(\mathbf{k},j)=\frac{1}{2}\left[\epsilon_{\mathbf{k}\sigma}+(-1)^{j}\sqrt{\epsilon_{\mathbf{k}\sigma}^{2}
-2z_{\sigma}J\epsilon_{\mathbf{k}\sigma}M+J^{2}\left<S^{2}\right>}\right]
\end{equation}\\
\noindent This way of determining the spectral function and therefore the Green's function is known 
as the spectral density approach(SDA)$^{15}$.  In this method, the only approximation is making an 
ansatz for $S_{\mathbf{k}\sigma}(E)$.  The rest of the calculation is exact.  Therefore, the method is nonperturbative
and is known to be more reliable in the strong coupling limit for the Hubbard model and Anderson model$^{15,16}$. In materials that we are aiming to study also, it is the strong coupling limit since $J$ is larger than the bandwidth.
Having determined $S_{\mathbf{k}\sigma}(E)$, the equation for the single-electron Green's
function becomes 
\begin{equation}
G_{\mathbf{k}\sigma}(E)=\left<\left<c_{\mathbf{k}\sigma};c_{\mathbf{k}\sigma}^{\dag}\right>\right>_{E}=
\sum_{j=1}^{2}\frac{\alpha_{\sigma}(\mathbf{k},j)}{E-E_{\sigma}(\mathbf{k},j)}
\end{equation}
Using the spectral theorem, the spin dependent occupation number is given by
\begin{equation}
n_{\sigma} = \frac {1} {N}\sum_{\mathbf{k}}\sum_{j=1}^{2}\left( \frac{\alpha_{\sigma}(\mathbf{k},j)}{e^{\beta(E_{\sigma}(\mathbf{k},j)-\mu)}+1}\right)
\end{equation}
Here $\beta=1/k_{B}T$. The chemical potential $\mu$ is fixed by the constraint
\begin{equation}
n=\sum_{\sigma}n_{\sigma}=constant.
\end{equation}
For the numerical evaluation of occupation numbers, the $\mathbf{k}$-summation can be conveniently
replaced by an integration over energy using a simple cubic (SC) density of states in a tight-binding approximation$^{20}$. In the next section we present the calculation of the static magnetic susceptibility.

\section {Static magnetic susceptibility}

\noindent In order to derive the magnetic phase diagram, one has to find out the magnetic ordering temperature $T_c$, which we find from the zeroes of the inverse paramagnetic susceptibility. The zero field static susceptibility of the itinerant electron subsystem is given by
\begin{equation}
  \chi(T)= \mu_{B} \sum_{\sigma}z_{\sigma}\left(\frac{\partial}{\partial B}
    n_{\sigma}\right)\Bigg|_{B\rightarrow 0}
\end{equation}
The local moment magnetization $\langle S^{z} \rangle$ and itinerant electron polarization 
$(n_{\uparrow}-n_{\downarrow})$ are mutually coupled and therefore
they become critical for the same parameters, in particular, at the same
temperature T. Therefore in the critical region, we  can reasonably assume

\begin{equation}
 \langle S^{z} \rangle\bigg|_{T>T_{c}}^{B \rightarrow 0} \propto \langle
 n_{\uparrow}-n_{\downarrow} \rangle\bigg|_{T>T_{c}}^{B \rightarrow 0} 
\end{equation}
This implies:
 \begin{equation}
   \mu_{B}\left(\frac{\partial}{\partial B}\langle S^{z} \rangle\right)\Bigg|_{T>T_{c}}^{B \rightarrow 0}=\eta\cdot\chi(T)
 \end{equation}
The proportionality factor $\eta$ has to be still fixed. This we do later.  A straightforward derivation of the itinerant-electron susceptibility eventually leads to the following expression:
\begin{equation}
\chi(T)=2\mu_{B}^2\left(\frac{A(T)+B(T)}{1-2\eta J[B(T)+C(T)]}\right)
\end{equation}
Where $A(T)$, $B(T)$ and $C(T)$ are temperature dependent integrals which are given by
\begin{equation}
A(T) =\frac {-\beta}{4N}\sum_{\mathbf{k}}\sum_{j=1}^{2}\left[ \alpha^{2}(\mathbf{k},j)Cosh^{-2}\left (\beta(E(\mathbf{k},j)-\mu)\right)\right] 
\end{equation}
\begin{equation}
B(T) =\frac {1}{N}\sum_{\mathbf{k}}\sum_{j=1}^{2}\left[(-1)^{j} \frac {J^{2}\left<S^{2}\right>} {2(\epsilon_{\mathbf{k}}^{2}+J^{2}\left<S^{2}\right>)^{3/2}}
\left( \frac{1}{e^{\beta(E(\mathbf{k},j)-\mu)}+1}\right)\right]
\end{equation}
\begin{equation}
C(T) =\frac {\beta}{4N}\sum_{\mathbf{k}}\sum_{j=1}^{2}\left[ (-1)^{j}\frac {\epsilon_{\mathbf{k}}} {2(\epsilon_{\mathbf{k}}^{2}+J^{2}\left<S^{2}\right>)^{1/2} } Cosh^{-2}\left (\beta(E(\mathbf{k},j)-\mu)\right)\right]
\end{equation}\\
Where  $\alpha(\mathbf{k},j)$ and $E(\mathbf{k},j)$ are the spectral weights and energies in the paramagnetic phase. The singularities are the solutions of the following equation:
\begin{equation}
0 = 1-2\eta J\Big[B(T)+C(T)\Big]\bigg|_{T=T_c}
\end{equation}
The instabilities of the paramagnetic phase towards ferromagnetism are thus given by the solutions of the above equation.

\begin{figure}[htb]
\begin{center}
    \epsfig{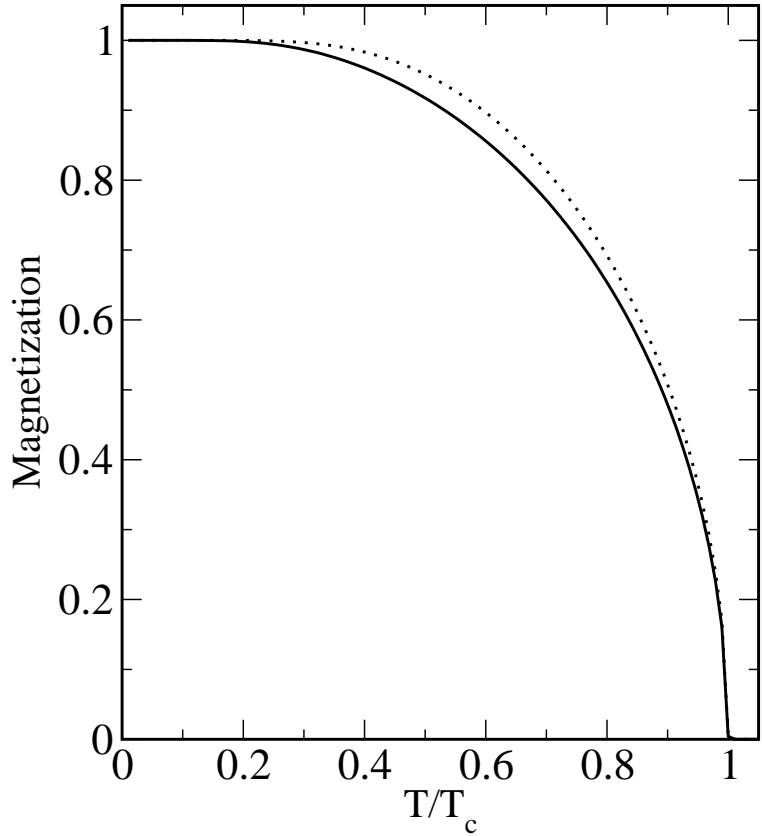}
    \caption{Local magnetization $\left<S^{z}\right>/S$ (full line) and itinerant electron polarization $( n_{\uparrow}-n_{\downarrow})/( n_{\uparrow}+n_{\downarrow})$(dotted line) with $n = 0.4$,  $S=3/2$ and $J=1$.}
\end{center}
\end{figure}

\begin{figure}[htb]
\begin{center}
    \epsfig{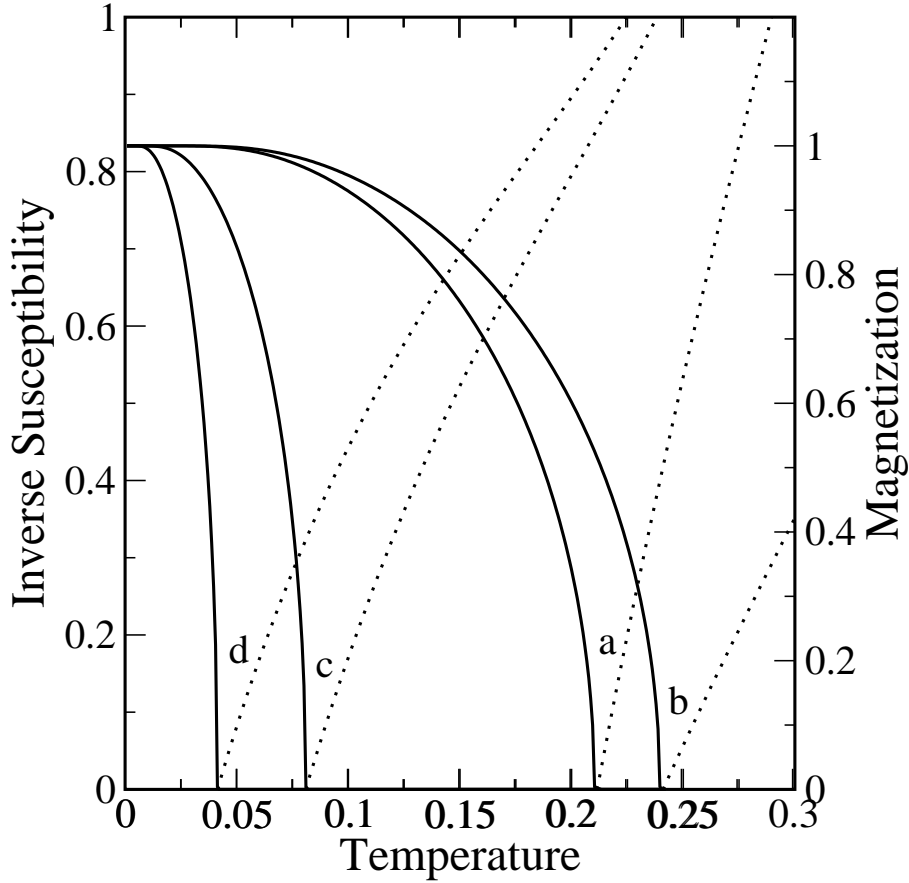}
    \caption{Temperature dependence of inverse susceptibility(dotted lines) and magnetization(full lines) for different values of
    band filling $n$. a) $n = 0.05$ b) $n = 0.10$ c) $n = 0.40$ and d) $n = 0.44$ with  $S=3/2$ and $J=1$.}
\end{center}
\end{figure}

\section {Results and discussion}

\noindent  For numerical evaluation we use SC density of states (DOS)$^{20}$ and the spin value $S=\frac{3}{2}$. We have chosen SC lattice density of states because it is one of the simplest with a peaked structure.  In contrast, a rectangular or semi-elliptic DOS would be, as it is known for a long time, very unfavourable to magnetism. $S=\frac{3}{2}$ represents the situation for manganites. The width W of the Bloch band has been chosen to be $1 eV$. Our theory does not aim at a self-consistent determination of the local moment magnetization  $M = <S^{z}>$ but rather at the influence of the interband exchange on the band states. It is known from the $D=\infty$ limit of KLM, that the thermal behaviour of $M$ is mean-field-like. We therefore consider $M$ as a parameter for which we choose a Brillouin function. We have fixed $\mu$ from the requirement given in Eq.(14). In order to fix the factor $\eta$ which is introduced in Eq.(17). We study the temperature dependence of local magnetization $\left<S^{z}\right>/S$ and itinerant electron polarization $( n_{\uparrow}-n_{\downarrow})/n$ in Fig.1. It is noted that  both the magnetizations behave in a similar fashion. The similar behaviour of both the magnetizations particularly in the critical region where transition takes place from magnetic to non-magnetic phase suggests the following analogy:
\begin{equation}
\frac{\langle S^{z}\rangle}{S}\Leftrightarrow \frac{( n_{\uparrow}-n_{\downarrow})}{n}
\end{equation}
So that we assume $\eta=S/n$. After fixing the value of $\mu$ for a given band occupancy $n$ and using this value of $\eta$, the paramagnetic susceptibility $\chi$ of the itinerant electron system given in Eq.(18) is calculated as a function of temperature. 

In manganites the interesting physics is due to the dynamics of the electrons in the d-band. This band is split by the crystal field into two, namely the triply degenerate $t_{2g}$-band and the doubly degenerate $e_g$-band. In the undoped situation, the $Mn$ is in the $Mn^{3+}$ and with four electrons in the d-band. The $t_{2g}$ band is occupied by three electrons with their spins aligned parallel due to strong Hund's rule coupling and this provides the localized spin $S=3/2$. The $e_g$-band is occupied by one electron. Upon doping sufficiently $Mn^{3+}$ state changes to $Mn^{4+}$ state with three d-electrons. Then only $t_{2g}$-band is occupied and the $e_g$ band is empty which means a hole is produced.  Since in the present calculation $n$ represents the number of itinerant electrons in the $e_g$-band, the cases with $n=1$ and $0$ represents the situation of $Mn^{3+}$ state and $Mn^{4+}$ state respectively. Therefore it is worthwhile to investigate the magnetic properties of the model system for given set of model parameters as a function of $n$. In order to investigate this, in Fig.2 we have plotted the inverse susceptibility as a function of temperature for various values of $n$. It is found that for various combinations of the model parameter values, the susceptibility follows Curie-Weiss behaviour. 

The magnetic ordering temperature $T_c$ is determined from the zeros of inverse susceptibility. The $n$ dependence of $T_c$ is shown in Fig.3. This figure clearly demonstrates that ferromagnetism does not exist if the band occupancy increases beyond a critical value  and this critical value itself increases with the increase of the interband exchange $J$. Further, for any choice of the values of model parameters $J$ and $W$, we did not find magnetism for $n>0.48$.  It is also noticed that the ferromagnetism is restricted to surprisingly low itinerant electron concentrations. It should be noted that the KLM used for the present investigation does not contain direct exchange between the localized moments. Therefore the collective ordering can only be mediated by conduction electron polarization. 

The steep increase of $T_c$ for low band occupations, a rather distinct maximum (which shifts to the lower values of $n$ with the increase of $J$) and then also a very rapid decrease to zero, qualitatively mimics the experimental observation on diluted ferromagnetic semiconductors such as $Ga_{1-x}Mn_xAs$. A similar result is found with the "modified" RKKY theory$^{21,22}$, interpolation self-energy ansatz method$^{13,23}$ and also in another theoretical work$^{24}$ using the dynamical mean field approximation. Though we did not inspect, we speculate that the break down of the ferromagnetic phase is followed by some kind of antiferromagnetism.

\begin{figure}[htb]
\begin{center}
    \epsfig{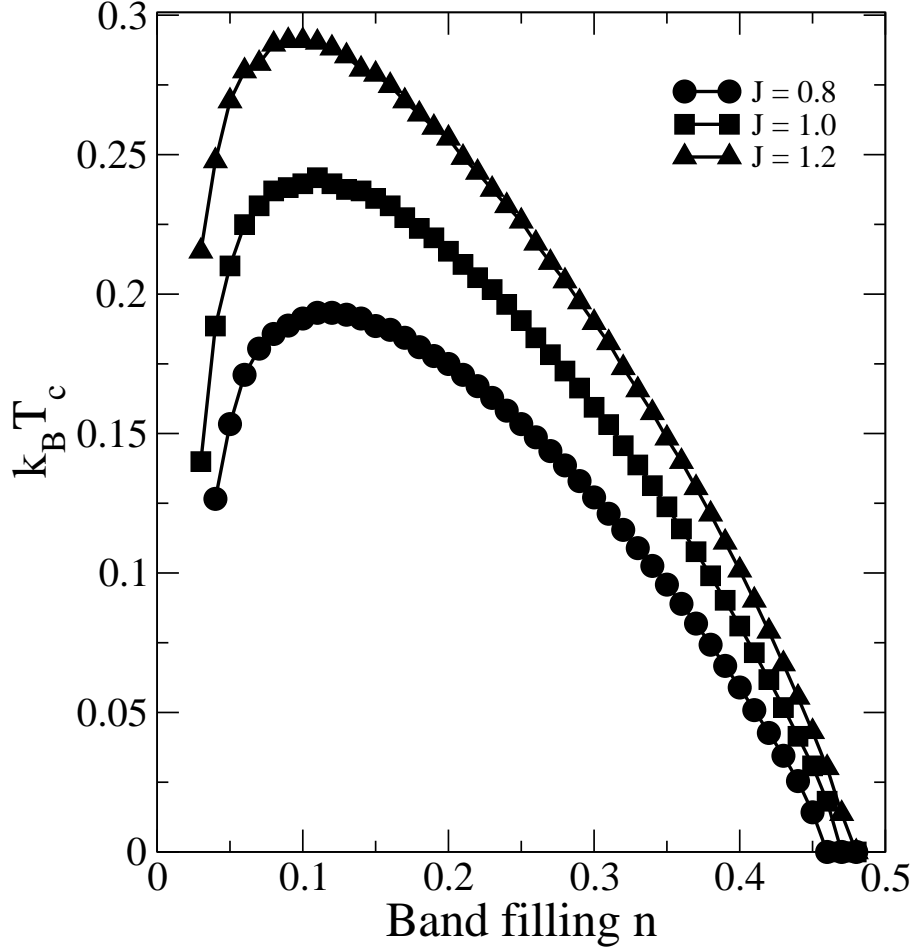}
    \caption{Curie temperature $T_c$ as a function of band filling $n$ for various values of $J$ and $S=3/2$.}
\end{center}
\end{figure}

\begin{figure}[htb]
\begin{center}
    \epsfig{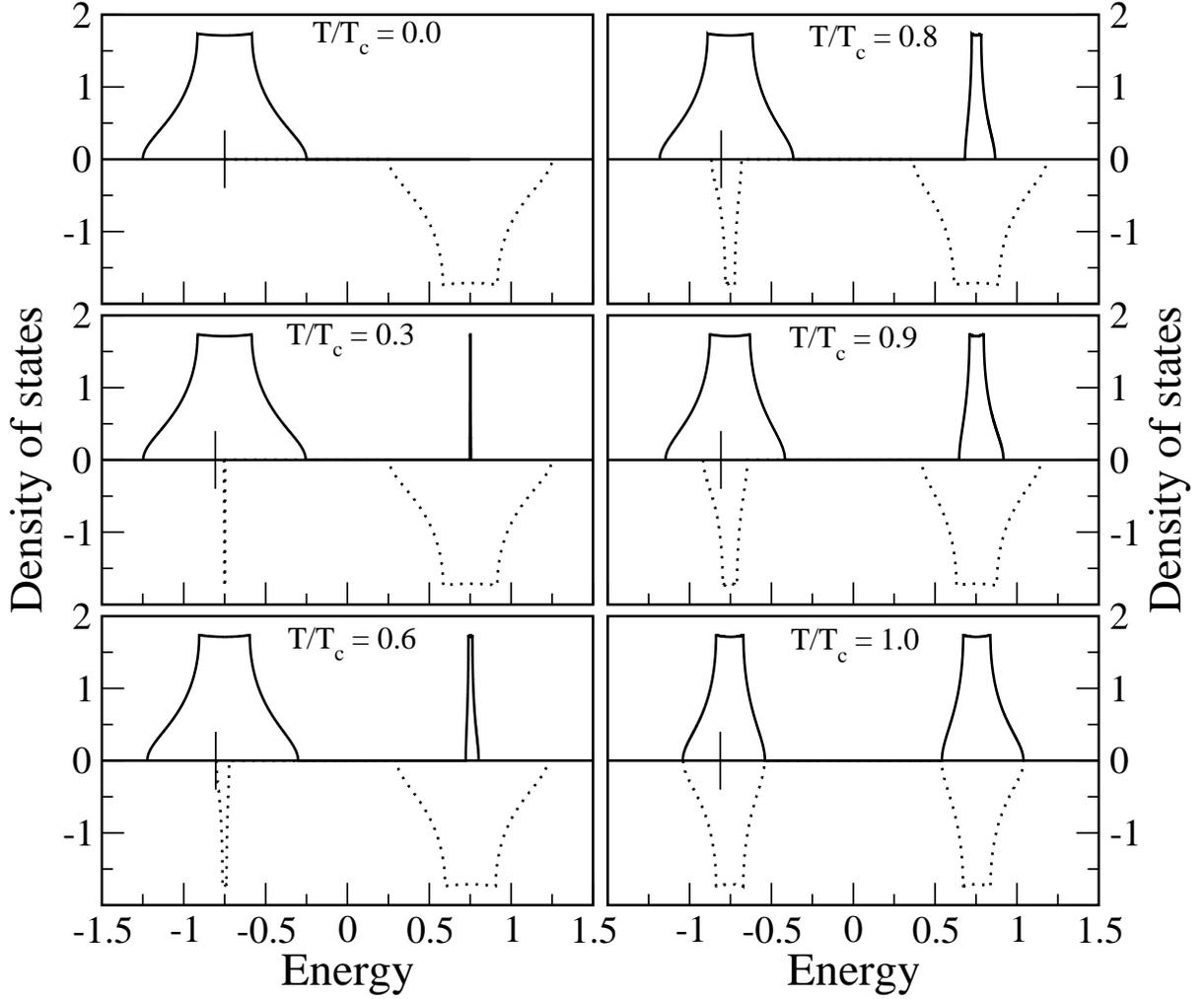}
    \caption{Quasiparticle density of states as a function of energy for various values of temperature.  
            Full line for up-spin and dotted line for down-spin. 
            The chemical potential is indicated by thin vertical line. Parameters: $n = 0.40$, $S=3/2$ and $J=1$.}
\end{center}
\end{figure}

Having determined the value of $T_c$, it is worthwhile to examine, particularly in the ferromagnetic region, the quasiparticle density of states (QDOS)  which we obtain from the single particle Greens function given in Eq.(12) through the relation. 

\begin{equation}
\rho_{\sigma}(E)= -\frac{1}{\pi N}\sum_{\mathbf{k}}\Im G_{\mathbf{k}\sigma}(E+i0^{+})
\end{equation}
The QDOS for various temperatures are displayed in Fig.4.

As is to be expected, the $e_g$-band splits for each spin direction into two sub-bands  centred at $\pm JS$. In the limit of large $J$ the spectral weights and energies given in Eqns.(10) and (11) reduce to

\begin{equation}
\alpha_{\sigma}(\mathbf{k},j)=\frac{1}{2}\left[1-(-1)^{j}z_{\sigma}\frac{M}{S}\right]\left[1+(-1)^{j}\frac{\epsilon
_{\mathbf{k}}}{JS}\left(1+(-1)^{j}z_{\sigma}\frac{M}{S}\right)\right]
\end{equation}

\begin{equation}
E_{\sigma}(\mathbf{k},j)=(-1)^{j}\frac{JS}{2}+\frac{\epsilon_{\mathbf{k}}}{2}\left[1-(-1)^{j}z_{\sigma}\frac{M}{S}\right]
\end{equation}

The separation of the bands and the spectral weights of these sub-bands however depend on $T$ through the value of $M$. For example, at $T=0$ ($M=S$), the spectral weight of the upper sub-band for $\uparrow$-states is zero.  The reason for this is easy to understand.  At $T=0$, the local moment system is saturated,  Therefore, for an $\uparrow$-electron there is no chance to spin-flip by involving a corresponding spin-flip of the local moment system.  That means, at $T=0$, as far as the $\uparrow$-electron is concerned, only the Ising part of $H_{sf}$ operates resulting simply in a rigid shift of band without any splitting.  The spectral weight of the $\downarrow$-states in the lower sub-band is however not zero.  This is because, for a $\downarrow$-electron, even at $T=0$, spin-flip is possible.  Further more, when a $\downarrow$-electron flips its spin, it lands as an $\uparrow$-electron.  Therefore the QDOS of the $\downarrow$-electron should be in the same energy region as that of $\uparrow$-electron.  For a $\downarrow$-electron there is another possibility. It can have repeated magnon emission and absorption.  That is, in a sense, it propagates in the lattice dressed by a cloud of magnons.  This is a stable quasiparticle, which we call the magnetic polaron. That is why for the $\downarrow$-spin there is a lower sub-band even at $T=0$. However, at $T=0$ the width of the sub-band is infinitesimally small or the DOS is $\delta$-function like and therefore is not shown in the figure.  Obviously, at $T=0$ there is no possibility of magnetic polaron for $\uparrow$-electron.  As $T$ increases ($M$decreases from saturation), the spin flip processes are allowed for both spin directions and therefore the spectral weights in both sub-bands are non-zero for both spin directions. At $T=T_C$ ($M=0$), the spectral weights of $\uparrow$- and $\downarrow$-states in the two sub-bands become equal as it should be. We note the asymmetry with respect to the center of the free band. This originates from the renormalization of the atomic levels by the $H_{sf}$ interaction$^{25}$. All these features
can be understood from the QDOS for various values of the temperature. In plotting this we use the magnetization obtained from the Brillouin function for a given band occupancy. 

Further, it can be noted that at $T=0$, for the lower sub-band, the width of the band, for $\downarrow$-states is zero and for $\uparrow$-states has twice the width of the free Bloch band and $\mu$ lies in the lower sub-band for $n=0.4$.  Therefore, the QDOS at $\mu$ has contribution only from $\uparrow$-electrons. This situation in literature is described$^{26-28}$ as a half-metal. The existence of half-metallic ferromagnets was first introduced by de Groot et al. In the case of doped manganites, half-metallicity was experimentally studied using low temperature resistivity data$^{27}$ and by analysing the spin-resolved photo emission data$^{29}$.  As the temperature is increased, the magnetization is reduced and the width of the band for states for $\uparrow$-electrons decreases whereas for states with $\downarrow$-electrons increases but the QDOS has contribution only from $\uparrow$-states.  When the temperature increases beyond certain critical value $T_h$ (which we call the transition temperature from half-metal to metal), the width of the band for $\downarrow$-states becomes sufficiently large so that the QDOS is contributed by $\downarrow$-states also. Thus one can see a transition from half-metal to metal as the temperature is varied. The critical temperature $T_h$ naturally should and does depend on $n$. For further increase of temperature, beyond $T_h$ the system undergoes transition from ferromagnetic to paramagnetic phase. It is found that $T_h$ depends on  $T_c$ and is less than $T_c$ which is consistent with the experimental findings from spin resolved photo emission studies$^{3}$ on $La_{0.7}Sr_{0.3}MnO_{3}$. Since $T_h$ and $T_c$ both depend on $n$ they can be varied by controlling the doping.  In addition since DOS depend on $M$, they can be manipulated by changing the external magnetic field in place of temperature and thereby again $T_h$ and $T_c$ can be changed. Therefore one can drive the system from half-metal to metal by tuning the magnetic field or doping. Since the change in QDOS and the shift in $\mu$ are continuous, the transition half-metal to metal is continuous.

\section {Conclusions}

KLM is used to study the possibility of collective magnetism in local-moment systems.  SDA which is nonperturbative and is known to be reliable in the strong coupling limit for the Hubbard model and the periodic Anderson model is used to solve KLM.  Using this approach, single-particle Greens function and the occupation numbers are calculated for itinerant electrons in the $e_g$-band.  The zero field static susceptibility of the itinerant electrons which are exchange coupled to localized magnetic moment is derived. Then, by determining the magnetic ordering temperature from the singularities of the susceptibility, the magnetic phase diagram is constructed.  It is found that the collective magnetism is possible only for low density of itinerant electrons. The QDOS of the $e_g$-band is studied as a function of temperature which shows the possibility of half-metal to metal transition in the region where the magnetic order is present.  The self-energy obtained in this approach is real which is an artifact of the ansatz made for the spectral function.  Therefore the effect of quasiparticle damping is beyond the scope of the present investigation.

\section*{Acknowledgements}

\noindent The authors are grateful to the Council of Scientific and Industrial Research, New Delhi, India for financial support through Grant No.03(1152)/10/EMR-II of a scheme.

\section*{References}

\noindent 1. T.V. Ramakrishnan, \textit {J. Phys.: Condens. Matter} \textbf{19}, 125211 (2007).\\
\noindent 2. A.P. Ramirez, \textit {J. Phys.: Condens. Matter} \textbf{9},  8171 (1997).\\
\noindent 3. J.-H. Park, E. Vescovo, H.-J. Kim, C. Kwon, R. Ramesh, and T. Venkatesan, \textit {Nature} \textbf{392}, 794 (1998).\\
\noindent 4. H. Akai, \textit {Phys. Rev. Lett.} \textbf{81}, 3002 (1998).\\
\noindent 5. M. Horne, P. Strange, W.M. Temmerman, Z. Szotek, A. Svane and H. Winter, \textit {J. Phys.: Condens. Matter} \textbf{16}, 5061 (2004) .\\
\noindent 6. Velicky, S. Kirkpatrik and Ehrenreich, \textit {Phys. Rev.} \textbf{175}, 747  (1968).\\
\noindent 7. G. Bulk and R.J. Jelitto, \textit {Phys. Rev.} \textbf{B41}, 413 (1990).\\
\noindent 8. W. Nolting, S. Rex and S.M. Jaya, \textit {J. Phys.:Condens. Matter} \textbf{9},  1301 (1997).\\
\noindent 9. S. Yunoki, J.Hu, A.L. Malvezzi, A. Moreo, N. Furukawa and E. Dagotto, \textit {Phys. Rev. Lett.} \textbf{80}, 845 (1998).\\
\noindent 10. M.Y. Kagan, D. I. Khomskii and M.V. Mostovoy, \textit {European Physical Journal} \textbf{B 12}, 217 (1999). \\
\noindent 11. A. Chattopadhyay and A. J. Mills and S. Das Sarma, \textit {Phys. Rev.} \textbf{B64},  012416 (2001).\\
\noindent 12. C. Lin and A.J. Millis, \textit {Phys. Rev.} \textbf{B 72}, 245112 (2005).\\
\noindent 13. W. Nolting, G.G. Reddy, A. Ramakanth and D. Meyer, \textit {Phys. Rev.} \textbf{B64}, 155109 (2001).\\
\noindent 14. T. Schneider, M. H. Pedersen and J. J. Rodríguez-Nún˜ez, \textit {Z. Phys. B-Condens. Matter} \textbf{100}, 263 (1996).\\
\noindent 15. T. Herrmann and W. Nolting, \textit {J. Magn. Mater.} \textbf{170}, (1997) 253 (1997).\\
\noindent 16. D. Meyer, W. Nolting, G. G. Reddy and A. Ramakanth, \textit {Physica Stat. Solidi (b)} \textbf{208}, 473 (1998).\\
\noindent 17. Maciej Maska, \textit {Phys. Rev. B} \textbf {48}, 1160 (1993).\\
\noindent 18. A. Schwabe and W. Nolting, \textit {Phys. Rev. B} \textbf {80}, 214408 (2009).\\
\noindent 19. L. Haritha, G. Gangadhar Reddy and A. Ramakanth \textit {Physica B} \textbf{405}, 1701 (2010). \\
\noindent 20. R. Jellito, \textit {J. Phys. Chem. Solids} \textbf{30},  609 (1969). \\
\noindent 21. W. Nolting, S. Rex, and S. Mathi Jaya, \textit {J. Phys.: Condensed Matter} \textbf{9}, 1301  (1997).\\
\noindent 22. C. Santos and W. Nolting, \textit {Phys. Rev.} \textbf{B65}, 144419 (2001).\\
\noindent 23. L. Haritha, G. Gangadhar Reddy, A. Ramakanth and S.K. Ghatak, to be published.\\
\noindent 24. A. Chattopadhyay, S. Das Sarma and A. J. Mills, \textit {Phys. Rev. Lett.} \textbf{87}, 227202 (2001).\\
\noindent 25. W. Nolting and M. Matlak, \textit {Phys. Status Solidi (b) } \textbf{123},  155 (1984).\\
\noindent 26. R. A. de Groot, F.M. Muller, P.G. van Engen and K.H.J. Buschow, \textit {Phys. Rev. Lett.} \textbf{50},  2024 (1983).\\
\noindent 27. V. Yu Irkhin and M. I. Katsnel'son, \textit {Physics Uspekhi} \textbf{37}, 659 (1994).\\
\noindent 28. G.- M. Zhao and H. Keller, W. Prellier and D.J. Kang, \textit {Phys. Rev.} \textbf{B63},  172411 (2001).\\
\noindent 29. J.-H. Park, E. Vescovo, H.-J. Kim, C. Kwon, R. Ramesh, and T. Venkatesan, \textit {Phys. Rev. Lett.} \textbf{81}, 1953 (1998).\\

\end{document}